\documentclass[aps,prd,twocolumn,superscriptaddress,nofootinbib]{revtex4-1}
\usepackage[utf8]{inputenc}
\usepackage[colorlinks=true,linkcolor=cyan,citecolor=cyan,urlcolor=cyan]{hyperref}
\usepackage{bm,bbm,amssymb,amsmath}
\usepackage{graphicx}
\usepackage[dvipsnames]{xcolor}

\newcommand{\Msun}{\ensuremath{\,M_{\odot}}}

\newcommand{\FZU}{CEICO, FZU-Institute of Physics of the Czech Academy of Sciences,
Na Slovance 1999/2, 182 21 Prague 8, Czech Republic}
\newcommand{\CITA}{Canadian Institute for Theoretical Astrophysics, University of Toronto, Toronto, Ontario M5S 3H8, Canada}

\begin{document}

\title{Prospects for constraining twin stars with next-generation \\ gravitational-wave detectors}

\author{Philippe Landry}
\affiliation{\CITA}
\email{plandry@cita.utoronto.ca}
\author{Kabir Chakravarti}
\affiliation{\FZU}
\email{chakravarti@fzu.cz}

\begin{abstract}
Neutron star equations of state with strong phase transitions may support twin stars, hybrid and hadronic stars with the same mass but different tidal deformabilities.
The presence of twin stars in the population of merging neutron stars produces distinctive gaps in the joint distribution of binary tidal deformabilities and chirp masses.
We analyze a simulated population of binary neutron star mergers recovered with a network of next-generation (XG) ground-based gravitational-wave detectors to determine how many observations are needed to infer, or rule out, the existence of twin stars. Using a hierarchical inference framework based on a simple parametric twin-star model, we find that a single week of XG observations may suffice to detect a tidal deformability difference of several hundred between twins and measure the mass scale at which twins occur to within a few percent. For less pronounced twins, XG observations will place a stringent upper bound on the tidal deformability difference.
\end{abstract}

\maketitle

\section{Introduction}

The phase structure of matter at the highest densities realized inside neutron stars is an unsolved puzzle for nuclear physics. Above the nuclear saturation density of $\rho_{\rm nuc} = 2.8\times 10^{14}$ g/cm$^3$, the nucleonic constituents of ordinary matter are thought to give way to other fundamental degrees of freedom, such as hyperons or deconfined quarks. If this phase transition occurs at a density that prevails inside neutron stars, then the heaviest of these compact objects may in fact be hybrid stars with exotic-matter cores. The possible existence of a stable family of hybrid stars with central densities greater than those of conventional hadronic neutron stars has been envisioned and studied extensively from the theoretical point of view~\cite{Gerlach1968, Kampfer1981, GlendenningKettner2000, SchertlerGreiner2000, AlfordHan2013, ZdunikHaensel2013, BenicBlaschke2015, AlfordBurgio2015, HanSteiner2019}.

In many nuclear theory models, the transition to the high-density phase is of first order, exhibiting a discontinuous first derivative of the baryon chemical potential with respect to the baryon density $\rho$. The hadronic- and exotic-matter phases may either remain separate and interface directly, or they may be linked by an intermediate mixed phase. In the former scenario, the so-called Maxwell construction, the baryon density experiences a discontinuity $\Delta\rho$ at fixed pressure $p$.  
If $\Delta\rho$ is sufficiently large, the dense matter equation of state may support stable twin stars: hadronic- and hybrid-star counterparts with the same mass, but different compactness~\cite{Schaffner-BielichHanauske2002, ZacchiTolos2017, AlfordSedrakian2017, ChristianZacchi2018}.

\begin{figure}[t]
    \includegraphics[width=0.95\columnwidth,trim={10 0 10 40},clip]{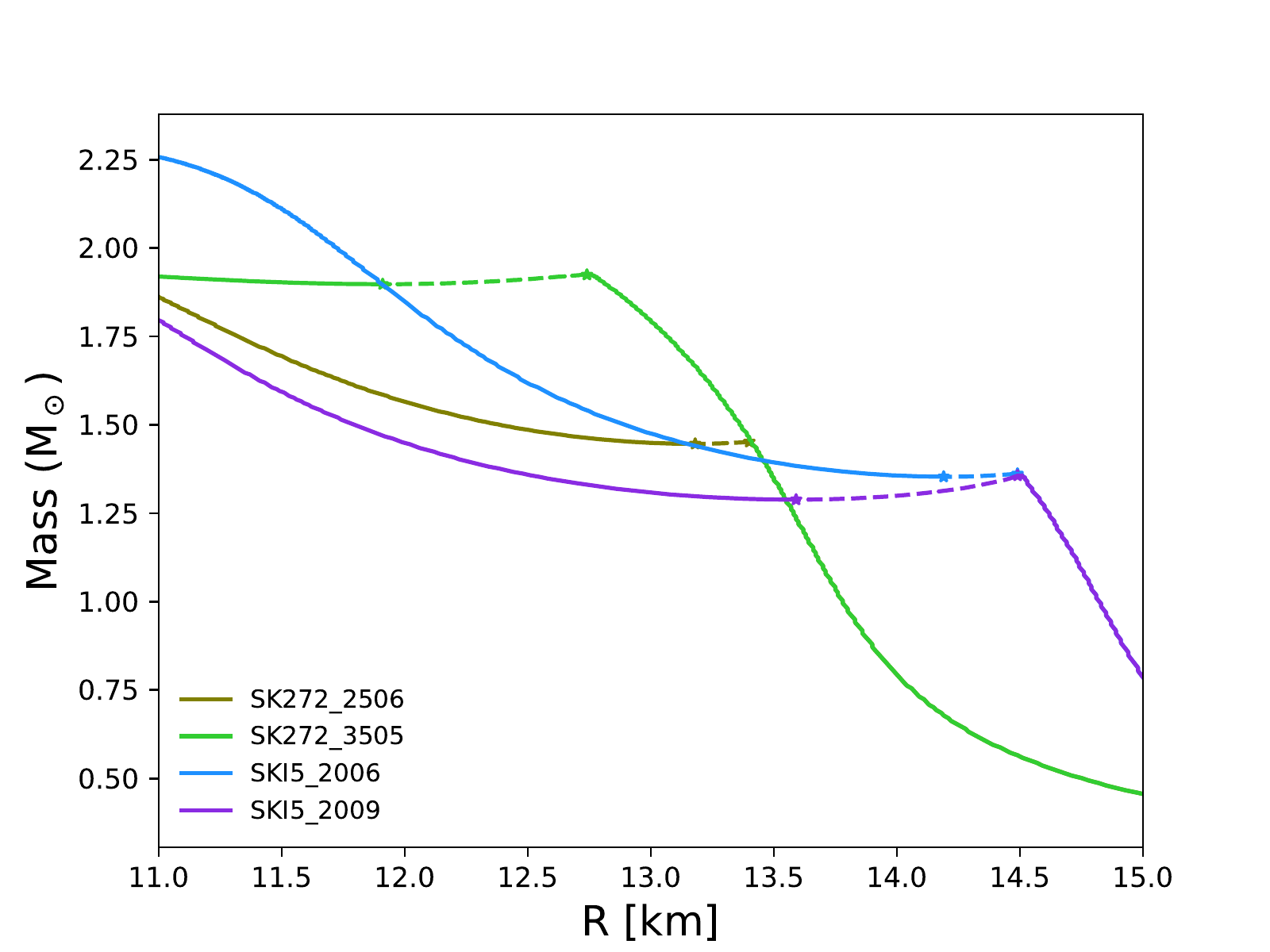} \\ 
    \includegraphics[width=0.95\columnwidth,trim={10 0 10 30},clip]{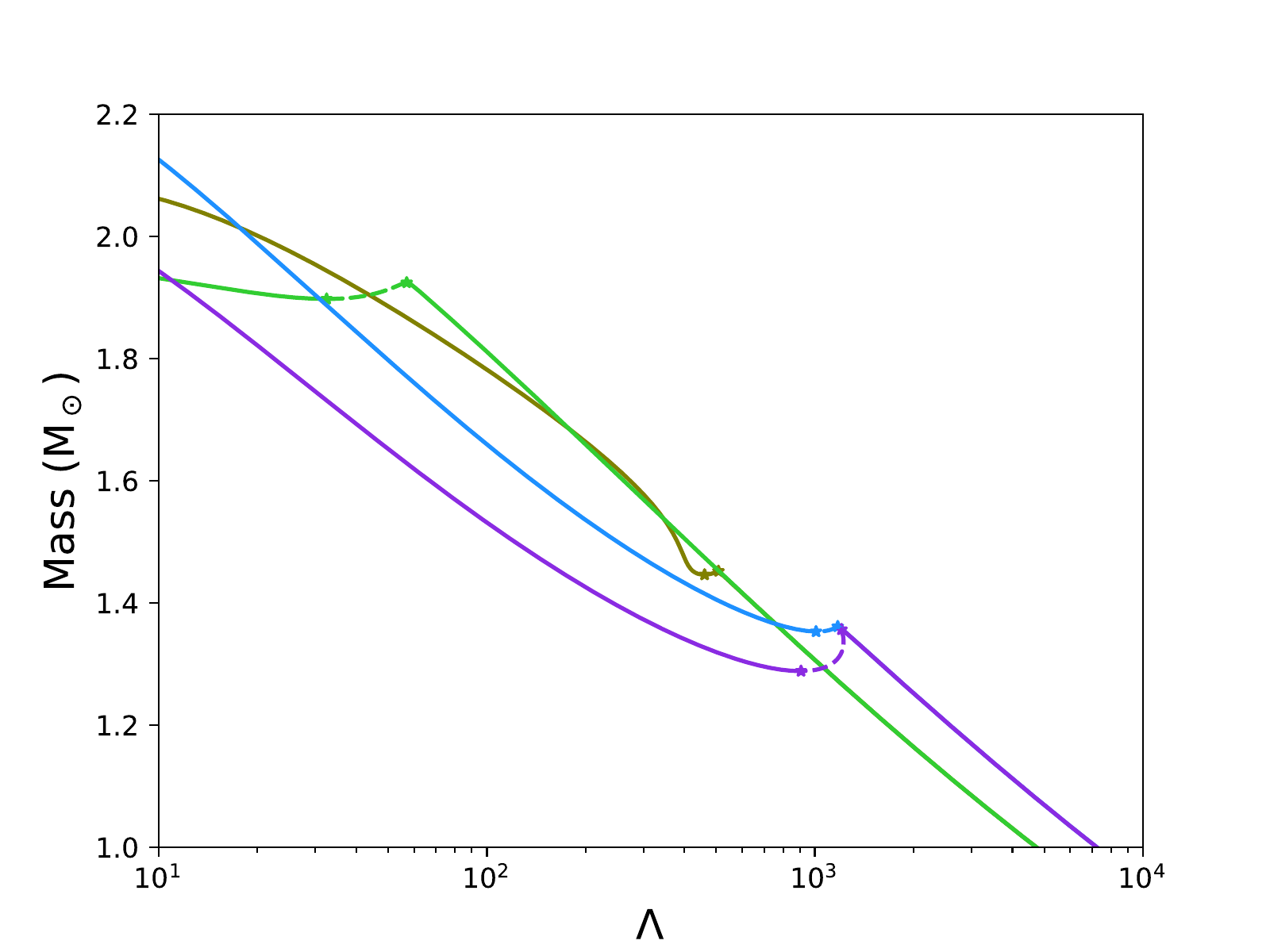}
    \caption{Sequences of neutron star properties for selected twin-star-supporting equations of state. We plot the mass-radius (top) and mass-tidal deformability (bottom) relations. Dashed lines indicate unstable segments of the sequence, where no neutron stars exist. Markers bookend the mass range over which twins occur.}
    \label{fig:twins}
\end{figure}

These twin-star configurations are visible in the mass-radius ($m$--$R$) relations of Fig.~\ref{fig:twins}, obtained by solving the Tolman-Oppenheimer-Volkoff (TOV) equations~\cite{OppenheimerVolkoff1939,Tolman1939} for selected equations of state. In each relation, there is a small range of masses for which a hybrid twin with a smaller radius and larger central density coexists alongside a hadronic star. Similar morphology is visible in the mass-tidal deformability ($m$--$\Lambda$) relations obtained by solving for a quadrupolar tidal perturbation~\cite{FlanaganHinderer2008,Hinderer2008,LandryPoisson2014}. The presence of radius or tidal deformability twins is a smoking gun for the occurrence of a strong first-order phase transition in neutron star matter. 

Because twin stars provide such a clear signature in the $m$--$R$ (or $m$--$\Lambda$) relation, their compatibility with existing observations of neutron stars has been heavily scrutinized. References~\cite{PaschalidisYagi2018,MontanaTolos2019,ChristianZacchi2019,PangDietrich2020,WangShi2022} demonstrated that parametric equations of state supporting twin stars could be constructed to simultaneously satisfy observational constraints on the tidal deformability from the compact binary merger GW170817~\cite{GW170817,LVC_GW170817source,LVC_GW170817eos} and bounds on the maximum mass $M_{\rm TOV}$ from the heaviest pulsars discovered in radio surveys~\cite{AntoniadisFreire2013,CromartieFonseca2020,FonsecaCromartie2021}. Reference~\cite{ChristianSchaffner-Bielich2020} argued that the radius measured for PSR J0030+0451 via X-ray pulse profile modeling~\cite{RileyWatts2019,MillerLamb2019} rules out twin stars associated with a phase transition onset density below $1.7\rho_{\rm nuc}$. Twin stars were also shown~\cite{ChristianSchaffner-Bielich2022} to be compatible with the inferred radius of PSR J0740+6620~\cite{RileyWatts2021,MillerLamb2021} and~\cite{LiSedrakian2021} with PREX-II's measurement of the neutron skin of $^{208}$Pb~\cite{AdhikariAlbataineh2021}. Meanwhile, Ref.~\cite{ChristianSchaffner-Bielich2021_Mmax} suggested (contra Ref.~\cite{TsaloukidisKoliogiannis2022}) that a maximum mass in excess of $2.2\Msun$ would exclude twin stars across the entire neutron star mass spectrum. Hence, although existing astronomical observations substantially restrict the parameter space for twin stars, they remain a live possibility.

\begin{figure}[t]
    \centering
    \includegraphics[width=0.95\columnwidth]{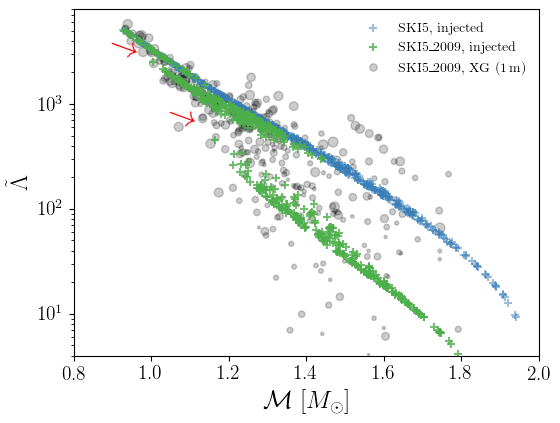}
    \caption{Joint binary tidal deformability and chirp mass distribution probed by one month of XG observations for the twin-star-supporting SKI5\_2009 equation of state, as compared with the purely hadronic SKI5 equation of state. A uniform neutron star mass distribution is assumed. Note the gaps in the distribution for SKI5\_2009 indicated by the arrows, which signify the existence of twin stars, compared to the contiguous SKI5 distribution. The detected distribution is subject to statistical uncertainties and is scattered by detector noise; for SKI5\_2009, we show the recovered chirp masses and binary tidal deformabilities in grey, with marker sizes inversely proportional to the signal-to-noise ratio. Actual statistical uncertainties are $O(10^2)$ in $\tilde{\Lambda}$ and $O(10^{-3})$ in $\mathcal{M}$.}
    \label{fig:lambdapop}
\end{figure}

Additional observations of neutron stars with the existing gravitational-wave detector network of Advanced LIGO~\cite{aLIGO}, Virgo~\cite{aVirgo} and KAGRA~\cite{AkutsuAndo2021} could, in principle, turn up a serendipitous discovery of twin stars. However, this is unlikely given the typically narrow mass range over which twins are supported and the high resolution in $\Lambda$ required to identify them---not to mention the modest expected binary neutron star detection rate~\cite{AbbottAbbott2018_ObservingScenarios,ColomboSalafia2022,PatricelliBernardini2022}.
Prospects for detecting twin stars with next-generation (XG) ground-based gravitational-wave observatories like Cosmic Explorer~\cite{EvansAdhikari2021} and Einstein Telescope~\cite{MaggioreVanDenBroeck2020} are more promising, thanks to their ability to capture virtually the complete merging neutron star population in the nearby Universe~\cite{EvansAdhikari2021,BorhanianSathyaprakash2022}. Although precisely measuring the tidal deformability of individual neutron stars will remain a challenging task,\footnote{Ref.~\cite{SmithBorhanian2021} reports statistical uncertainties of $O(100\%)$ in component tidal deformabilities $\Lambda_{1,2}$ even for a simulated XG binary neutron star merger at 40 Mpc with signal-to-noise ratio 2400.} twin-star-supporting equations of state give rise to a distinctive distribution of binary tidal deformabilities $\tilde{\Lambda}$ vs chirp masses $\mathcal{M}$ across the population~\cite{ChatziioannouHan2020}: as shown in Fig.~\ref{fig:lambdapop}, when no twin stars are present, the distribution is contiguous, while twins introduce gaps that make it disjoint. This population-level signature can be used to infer the existence of twin stars even when individual twin-star pairs are not identifiable. Here we show that XG measurements of the binary tidal deformability distribution can detect the existence of twin stars at 90\% confidence (or with a log Bayes factor greater than 6) with as few as $\sim 100$ observations.

To do so, we first construct a family of equations of state with first-order phase transitions giving rise to stable twin stars, and simulate populations of binary neutron star mergers recovered with different gravitational-wave detector networks. We then collect the binary tidal deformability and chirp mass measurements from each population. Heuristically, twin stars are detected at the population level when the recovered distribution is disjoint, rather than contiguous. To implement this test systematically, we develop a Bayesian hierarchical inference framework with a parametric model for the $m$--$\Lambda$ relation, formulated in terms of a twin-star mass scale $M_t$ and the tidal deformability difference $\Delta\Lambda$ between twins. This approach follows Ref.~\cite{ChatziioannouHan2020}, which tackled the related problem of inferring the radius difference between hybrid and hadronic stars with LIGO and Virgo.
We find that current gravitational-wave detectors, even at upgraded ``A+'' sensitivities~\cite{AbbottAbbott2018_ObservingScenarios}, cannot meaningfully constrain the twin-star parameters. However, in the best-case scenario we consider, an XG network can determine $M_t$ and $\Delta\Lambda$ to within 1\% and 15\%, respectively, at 90\% confidence after just one month of observations.

\section{Twin-star equations of state}\label{Sec_PT}

We construct equations of state giving rise to twin stars using the constant-sound-speed formulation of Ref.~\cite{HanSteiner2019}. It consists of a low-density hadronic equation of state, to which a constant-pressure first-order phase transition segment of length (or ``strength'') $\Delta\rho$ is connected at onset density $\rho_t$. A high-density extension with constant sound speed is appended to the end of the phase transition segment.

We demand that the equations of state satisfy existing constraints from neutron star observations. If we suppose that the $\sim 1.4\Msun$ neutron stars observed in GW170817 and as PSR J0030+0451 are purely hadronic, then the low-density equation of state must satisfy the approximate observational upper bounds $\Lambda_{1.4} \lesssim 580$~\cite{LVC_GW170817eos} and $R_{1.4} \lesssim 14$ km~\cite{MillerLamb2019}.\footnote{In fact, because the analysis in Ref.~\cite{LVC_GW170817eos} relies on equation-of-state-insensitive relations that can't accommodate twin stars, we allow for some leeway in the first constraint.} If, instead, we suppose that those compact objects are hybrid stars, then the low-density portion of the equation of state remains unconstrained. These considerations motivate the choice of SKI272 ($R_{1.4} \approx 13.5$ km, $\Lambda_{1.4} \approx 650$)~\cite{AgrawalShlomo2003} and SKI5 ($R_{1.4} \approx 14.5$ km, $\Lambda_{1.4} \approx 1000$)~\cite{ReinhardFlocard1995,BertulaniValencia2019}, respectively. Both of these low-density equations of state have a symmetry energy slope $L \lesssim 140$, in keeping with the implications of the PREX-II experiment~\cite{ReedFattoyev2021}.

\begin{table*}[t]
    \centering
    \begin{tabular}{lcccccccc}
    \hline \hline
    EOS & $\rho_t\;[\rho_{\rm nuc}]$ & $\Delta\rho\;[\rho_{t}]$ & $\Delta M_{t}\;[M_{\odot}]$ & $M_{\rm TOV}\;[M_{\odot}]$ & $R_{1.4}\;$[km] & $\Lambda_{1.4}$ & $M_t\;[M_{\odot}]$ & $\Delta\Lambda$ \\ \hline    
     SKI5\_2006 & 2.0 & 0.6 & 0.01 & 2.28 & 13.5 & 520  & 1.36 & 340 \\
     SKI5\_2009 & 2.0 & 0.9 & 0.07 & 2.08 & 12.2 & 240 & 1.32 & 940 \\
     SK272\_2506 & 2.5 & 0.6 & 0.01 & 2.08 & 13.5 & 650 & 1.45 & 80 \\ 
     SK272\_3505 & 3.5 & 0.5 & 0.03 & 1.94 & 13.5 & 650 & 1.91 & 40 \\ \hline \hline
    \end{tabular}
    \caption{Twin-star-supporting equations of state used in this study. The parameters $\rho_t$ and $\Delta\rho$ specifying the onset density and strength of the first order phase transition are listed. The mass range $\Delta M_t$ over which twins are supported, the maximum mass $M_{\rm TOV}$, and the canonical radius $R_{1.4}$ and tidal deformability $\Lambda_{1.4}$ for the corresponding sequence of neutron stars are also given. Additionally, we list the twin-star mass scale $M_t$ and tidal deformability difference $\Delta\Lambda$ that we seek to recover via hierarchical inference. The equation of state's name indicates the low-density hadronic model upon which it is based.}
    \label{tab:eos}
\end{table*}

For the high-density extension, we select a causal equation of state to increase the odds that stable twin stars will be supported. To interpolate between the high- and low-density regimes, we explore different first-order phase transition segments by selecting various combinations of onset density $\rho_t \in [1, 4]\rho_{\rm nuc}$ and strength $\Delta\rho \in (0,2]\rho_t$. For use in our study, we retain a representative subset of those equations of state that give rise to stable twin stars while simultaneously satisfying the approximate mass and radius constraints from electromagnetic observations of PSR J0740+6620: $M_{\rm TOV} \gtrsim 2.0\Msun$, $R_{2.0} \gtrsim 12$ km~\cite{MillerLamb2021}. The selected equations of state, their phase transition parameters and their associated neutron star observables are listed in Table~\ref{tab:eos}. Their $m$--$R$ and $m$--$\Lambda$ relations are plotted in Fig.~\ref{fig:twins}. Of particular relevance for our analysis is the presence of an unstable segment in the $m$--$\Lambda$ relation that connects the hadronic and hybrid branches at a mass scale corresponding to $\rho_t$.

\section{Binary neutron star population and tidal deformability distribution}\label{Sec_BNSPop}
To determine when twin stars are identifiable in the binary neutron star population, we generate simulated distributions of masses and binary tidal deformabilities. We prescribe a neutron star mass model, select an equation of state, sample a realization of the astrophysical population of binary neutron star mergers, and determine which events are detected by computing their optimal signal-to-noise ratio with respect to the detector networks of interest.

Inspired by studies of the gravitational-wave population to date~\cite{LandryRead2021,LVK_O3bPop}, our fiducial neutron star mass distribution is uniform for $m \in [1\Msun,M_{\rm TOV}]$. 
We assume that both components of a neutron star binary are drawn from this common mass distribution, and that they pair randomly.
We distribute the sources isotropically on the sky and according to a Madau-Dickinson star formation rate~\cite{MadauDickinson2014} in redshift. 
We adopt a local, astrophysical binary neutron star merger rate of 440 Gpc$^{-3}$ y$^{-1}$, the 90\% confidence upper bound from Ref.~\cite{LVK_O3bPop}'s PowerLaw+Dip+Break model.
The binary tidal deformability for each event is determined from its masses by the $m$--$\Lambda$ relation dictated by the equation of state. Neutron stars in the twin-star mass range are randomly assigned to the hadronic or the hybrid branch of the relation with equal probability. 

We calculate optimal signal-to-noise ratios for the simulated mergers with respect to three different detector networks: the ``HLV'' network includes LIGO-Hanford, LIGO-Livingston and Virgo detectors at their design sensitivities; the ``A+'' network upgrades the LIGO-Hanford and LIGO-Livingston detectors to A+ sensitivity; and the ``XG'' network replaces the LIGO detectors with Cosmic Explorer detectors, and the Virgo detector with Einstein Telescope, at their respective design sensitivities. For the HLV and A+ networks, we simulate mergers within their projected luminosity distance ranges of 190 Mpc and 330 Mpc, respectively, for binary neutron stars~\cite{AbbottAbbott2018_ObservingScenarios}. 
For the XG network, whose binary neutron star range will extend to cosmological distances~\cite{EvansAdhikari2021,BorhanianSathyaprakash2022}, we simulate only the nearby mergers ($z \lesssim 0.5$) that contribute the loudest events. Our signal-to-noise ratio calculation is carried out with \texttt{bilby}~\cite{AshtonHubner2019}, using the \texttt{IMRPhenomPv2\_NRTidal} waveform model~\cite{DietrichKhan2019}, a minimum frequency of 10 Hz (40 Hz for HLV), and a network signal-to-noise ratio threshold of 12 for detection.

Given the assumed merger rate and distribution of sources, our simulated populations yield expected astrophysical rates of 3, 15 and 5800 binary neutron star mergers per year within the aforementioned ranges for HLV, A+ and XG, respectively. The exact number of detections depends on the population realization, but the detection efficiency is $\sim 60\%$ for A+ and $\sim 95\%$ for XG. Since the HLV network makes so few detections, we focus our analysis on the A+ and XG scenarios. For each detected event, we simulate the likelihoods in chirp mass $\mathcal{M}$, mass ratio $q$ and binary tidal deformability $\tilde{\Lambda}$ as independent Gaussians, scaling their standard deviations with the event's signal-to-noise ratio in inverse proportion to GW170817's~\cite{FarrBerry2016}. We mock up the effect of the detector noise realization by adding Gaussian noise that shifts the median of the likelihood within one standard deviation.

An example of the resulting observed distribution of binary tidal deformabilities vs chirp masses is illustrated in Fig.~\ref{fig:lambdapop} for SKI5\_2009. One can clearly see the disjoint nature of the true, underlying distribution, but it is less visible when the data is scattered by detector noise. For equations of state with smaller $\Delta\Lambda$, the gaps in the distribution may not be discernible by eye at all. 
The need to simultaneously account for statistical uncertainty in the masses and tidal deformabilities of individual events, as well as uncertainty in the underlying $m$--$\Lambda$ relation, motivates the elaboration of the hierarchical inference framework we describe below.

\section{Hierarchical Inference}\label{Sec_inference}

We systematically search for evidence of twin stars in the binary tidal deformability vs chirp mass distribution using a simple model. We prescribe a parametric population model

\begin{equation}
    \pi(\mathcal{M},q,\tilde{\Lambda}|\boldsymbol{\theta},\boldsymbol{\lambda}) = \pi(\tilde{\Lambda}|\mathcal{M},q;\boldsymbol{\theta}) \pi(\mathcal{M},q|\boldsymbol{\lambda})
\end{equation}
built from a binary tidal deformability distribution

\begin{align} \label{lambdadist}
    \pi(\tilde{\Lambda}|\mathcal{M},q;\boldsymbol{\theta}) = \int& P(\tilde{\Lambda}|\Lambda_1,\Lambda_2,q) \pi(\Lambda_1|m_1;\boldsymbol{\theta}) \pi(\Lambda_2|m_2;\boldsymbol{\theta}) \nonumber \\ 
    &\times P(m_1,m_2|\mathcal{M},q) d\Lambda_1 d\Lambda_2 dm_1 dm_2 
\end{align}
that depends on parameters $\boldsymbol{\theta}$ describing an $m$--$\Lambda$ relation $\pi(\Lambda|m;\boldsymbol{\theta}) = \delta(\Lambda - \Lambda_{\boldsymbol{\theta}}(m))$, and a mass distribution

\begin{equation}
    \pi(\mathcal{M},q;\boldsymbol{\lambda}) = \int P(\mathcal{M},q|m_1,m_2) \pi(m_1,m_2|\boldsymbol{\lambda}) dm_1 dm_2
\end{equation}
that depends on parameters $\boldsymbol{\lambda}$, expressed here in terms of a joint distribution $\pi(m_1,m_2|\boldsymbol{\lambda})$ for the component masses $m_{1,2}$. A hierarchical inference of the population model parameters $\boldsymbol{\theta}, \boldsymbol{\lambda}$ proceeds according to~\cite{MandelFarr2019}

\begin{align} \label{bayes}
    P(\boldsymbol{d}|\boldsymbol{\theta},\boldsymbol{\lambda}) = \prod_i \frac{1}{\zeta(\boldsymbol{\theta},\boldsymbol{\lambda})} \int& P(d_i | \mathcal{M},q,\tilde{\Lambda}) \nonumber \\ 
    &\times \pi(\mathcal{M},q,\tilde{\Lambda} | \boldsymbol{\theta},\boldsymbol{\lambda}) \,d\mathcal{M}\, dq\, d\tilde{\Lambda} ,
\end{align}
where $P(d_i|\mathcal{M},q,\tilde{\Lambda})$ is the marginal gravitational-wave likelihood in chirp mass, mass ratio and binary tidal deformability for the $i$th observation, and $\zeta(\boldsymbol{\theta},\boldsymbol{\lambda})$ is the fraction of the population that is detected. Since we are only interested in the $m$--$\Lambda$ relation parameters $\boldsymbol{\theta}$ in our application, we fix the mass distribution as specified above. We also ignore selection effects, as they do not impact the recovery of the parameters of interest: in our population model, the detection fraction is approximately independent of the equation of state.\footnote{The gravitational-wave selection function for binary neutron star mergers is proportional to $\mathcal{M}^{5/2}$~\cite{ChatziioannouFarr2020}, so the detection fraction depends on the equation of state through the maximum mass in the population, set here by $M_{\rm TOV}$. However, $M_{\rm TOV}$ varies by only $\sim 10\%$ across the chosen equations of state.}
Thus, we approximate Eq.~\eqref{bayes} as
\begin{equation}
    P(\boldsymbol{d}|\boldsymbol{\theta}) \propto \prod_i \int P(d_i | \mathcal{M},q,\tilde{\Lambda}) \, \pi(\mathcal{M},q,\tilde{\Lambda} | \boldsymbol{\theta}) \, d\mathcal{M} \, dq \, d\tilde{\Lambda} .
\end{equation}
We adopt uniform priors on the $m$--$\Lambda$ relation parameters $\boldsymbol{\theta}$, such that the posterior $P(\boldsymbol{\theta}|\boldsymbol{d})$ is proportional to this equation.

To evaluate the likelihood $P(\boldsymbol{d}|\boldsymbol{\theta})$, we make a Monte Carlo approximation for the integrals over $\mathcal{M}$ and $q$. The remaining integral is resolved by the delta function in $\tilde{\Lambda}$ that results from Eq.~\eqref{lambdadist}. We then have

\begin{align}
    P&(\boldsymbol{d}|\boldsymbol{\theta}) \approx \nonumber \\ 
    &\prod_i \sum_j P(d_i | \mathcal{M}_j,q_j,\tilde{\Lambda}_{\boldsymbol{\theta}}(\mathcal{M}_j,q_j)) \; | \; \mathcal{M}_j, q_j \sim \pi(\mathcal{M},q)
\end{align}
up to an overall normalization. We sample from this likelihood using a Markov-chain Monte Carlo algorithm implemented with \texttt{emcee}~\cite{Foreman-MackeyHogg2013}.

To model the twin-star $m$--$\Lambda$ relations, we adopt the parametric form

\begin{figure}[h]
    \centering
    \includegraphics[width=0.95\columnwidth]{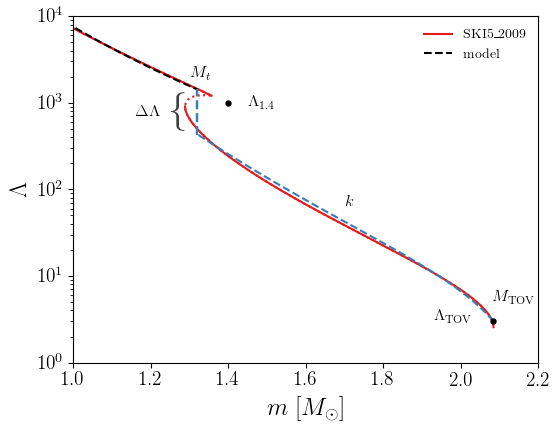}
    \caption{Parametric model for the mass-tidal deformability relation used in the hierarchical inference. The fit to the relation for the SKI5\_2009 equation of state is shown. The parameters we infer are highlighted in blue. The parameters we fix for simplicity are shown in black.}
    \label{fig:diagram}
\end{figure}

\begin{widetext}
\begin{equation}
    \Lambda_{\boldsymbol{\theta}} = \begin{cases} \Lambda_{1.4}^{\rm had} \left(m/1.4 \Msun\right)^{-6} & 1\Msun \leq m \leq M_t \\
    \left[\Lambda_{1.4}^{\rm had} \left(M_t/1.4 \Msun\right)^{-6} - \Delta\Lambda\right] \left[\left(\frac{m}{M_t}\right)^{-k} - \frac{(1-m/M_t)}{(1-M_{\rm TOV}/M_t)} \left(\frac{M_{\rm TOV}}{M_t}\right)^{-k}\right] +\Lambda_{\rm TOV} \frac{(1-m/M_t)}{(1-M_{\rm TOV}/M_t)} & M_t < m \leq M_{\rm TOV} \\
    \end{cases}
\label{fit}
\end{equation}
\end{widetext}
with parameters $\boldsymbol{\theta} = \lbrace \Lambda_{1.4}^{\rm had}, M_{t}, \Delta\Lambda, k, M_{\rm TOV}, \Lambda_{\rm TOV} \rbrace$---the tidal deformability of a canonical hadronic neutron star, the (median) twin-star mass, the tidal deformability difference between twins of mass $M_t$, the power-law slope of $\Lambda(m)$ on the hybrid branch, the maximum neutron star mass and the tidal deformability of the maximum-mass star, respectively. To keep the model simple, we do not model the finite extent of the twin star mass range; $M_{t}$ simply represents its midpoint. Thus, the twin-star signature that informs the hierarchical inference is the existence of a gap of width $\Delta\Lambda$ in the tidal deformability distribution at $M_{t}$, rather than the multimodality of the tidal deformability distribution itself. This model for the mass-tidal deformability relation is shown in Fig.~\ref{fig:diagram}.

The uniform priors for the $m$--$\Lambda$ relation parameters are subjected to the constraints $1.00\Msun \leq M_{t} \leq 1.94\Msun$, $0 \leq \Delta\Lambda \leq 1500$, $5 \leq k \leq 12$. Finding that $\Lambda_{1.4}$ is always well-recovered, and that the results are relatively insensitive to the maximum-mass parameters, we fix $M_{\rm TOV}$, $\Lambda_{1.4}$ and $\Lambda_{\rm TOV} = 3$ for simplicity.

\section{Results}\label{Sec_res}

We apply the hierarchical inference of twin-star parameters to the populations simulated for the hybrid equations of state listed in Table~\ref{tab:eos}, as well as the purely hadronic equations of state SKI5 and SK272. For the hadronic cases, we expect to recover the prior on $M_{t}$ and a posterior peaked at $\Delta\Lambda = 0$. For the hybrid cases supporting twin stars, we expect to recover a posterior on $\Delta\Lambda$ peaked away from zero, and a constraint on $M_{t}$. We consider the twin stars as detected if the marginal $\Delta\Lambda$ highest-posterior-density interval excludes zero at 90\% confidence. As an alternative quantification of the evidence for twin stars, we also compute an evidence ratio (Bayes factor) between the hypotheses that twin stars are present and absent in the population.

\begin{figure*}[t]
    \centering
    \includegraphics[width=0.95\textwidth,trim={145 160 130 70},clip]{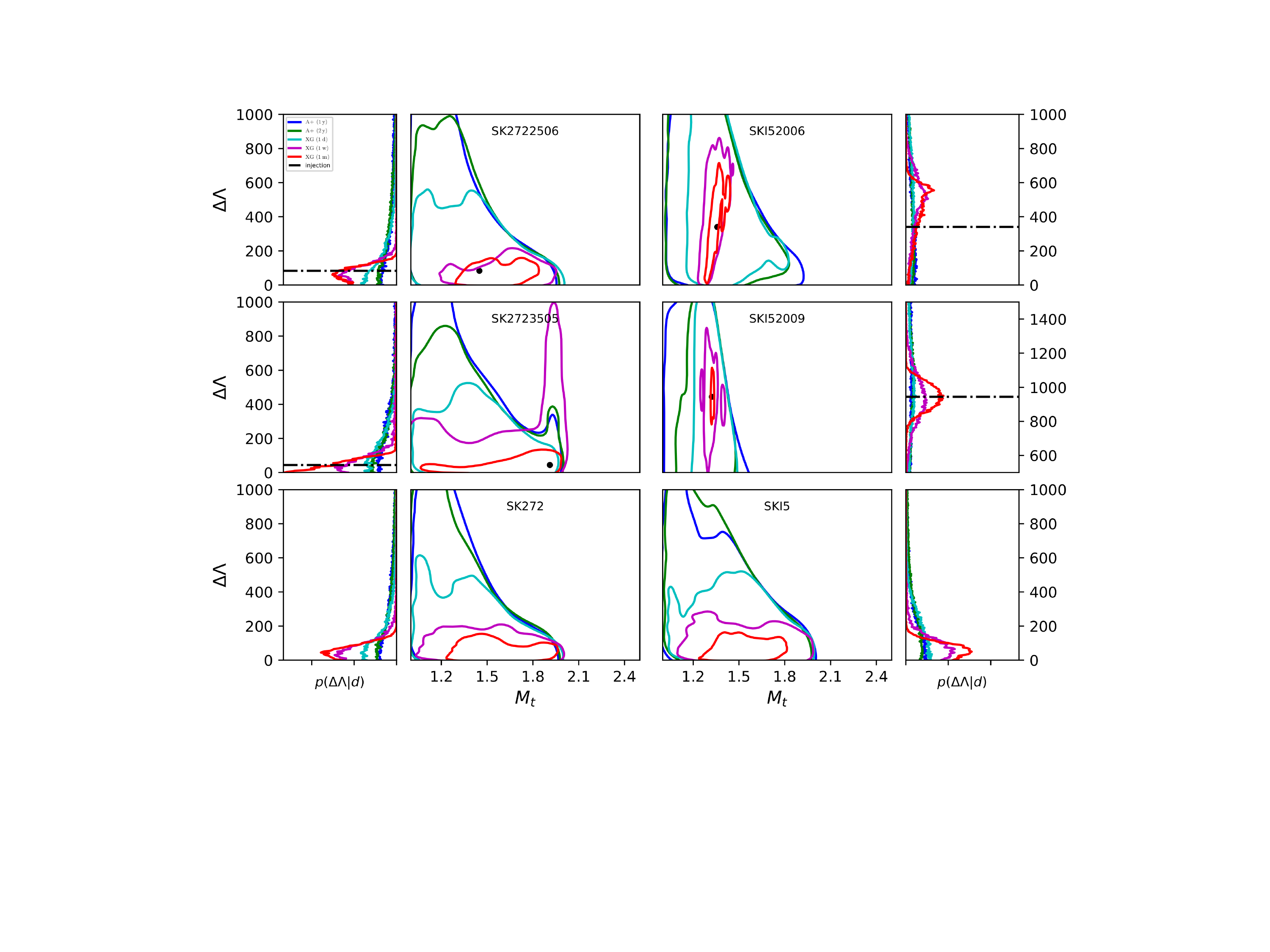}
    \caption{Posteriors on twin-star parameters of the mass-tidal deformability relation for different equations of state. The 90\% credible region of the posterior, averaged over 10 population realizations, is shown for various observing scenarios. The parameters corresponding to the injected equation of state are indicated in black; in the bottom two panels, where the injected equation of state is purely hadronic, the $M_t$ parameter has no intrinsic meaning.}
    \label{fig:results}
\end{figure*}

The recovered twin star parameters for each equation of state are shown in Fig.~\ref{fig:results} for a sequence of observing scenarios, from one year at A+ sensitivity to one month at XG sensitivity. We show the average recovery over 10 realizations of the population to mitigate statistical fluctuations in the confidence regions. 
In the case of the equation of state with the strongest phase transition, SKI5\_2009, the one-dimensional marginal posterior on $\Delta\Lambda$ already comfortably favors the presence of twin stars at 90\% confidence after one week of observation with the XG network. The $\Delta\Lambda$ posterior for SKI5\_2006, which has the same onset density but a somewhat weaker phase transition, also satisfies our criterion for twin-star detection after one week of XG observations.

For the equations of state with higher onset densities, and correspondingly smaller $\Delta M_t$ and $\Delta\Lambda$, the tidal deformability resolution in the XG observations is not sufficient to distinguish between zero and finite $\Delta\Lambda$. However, after one month of XG observations, $\Delta\Lambda$ larger than $\sim 100$ can be ruled out at 90\% confidence for both SK272\_2506 and SK272\_3505. This resolution limit in tidal deformability also holds for the recoveries of the hadronic equations of state SKI5 and SK272: the $\Delta\Lambda$ posteriors converge towards zero as expected, but cannot exclude $\Delta\Lambda \lesssim 100$.

Hence, based on the 90\% confidence intervals for $\Delta\Lambda$, we correctly conclude after one month of XG observations that twin stars are present in the binary neutron star populations for SKI5\_2009 and SKI5\_2006, and that twin stars with $\Delta\Lambda \gtrsim 100$ are excluded in the other four scenarios. In all cases, the $\Delta\Lambda$ posteriors after even two years of A+ observations are essentially uninformative. This demonstrates that an XG detector network is needed to definitively identify (or rule out) twin stars with gravitational waves at the population level.

\begin{figure}[t]
    \centering
    \includegraphics[width=0.95\columnwidth]{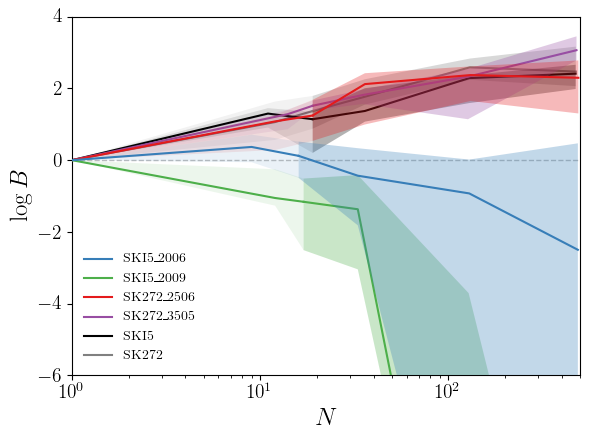}
    \caption{Model comparison between twin-star and no-twin hypotheses for each injected equation of state. We show how the log Bayes factor evolves as a function of the number of observations in the A+ era (light shading) and the XG era (dark shading). A negative $\log B$ favors the presence of twin stars. The shaded regions encompass 68\% confidence intervals on the Bayes factor due to its variation across 10 different population realizations.}
    \label{fig:bfs}
\end{figure}

The evidence ratios between twin and no-twin hypotheses tell a similar story. For each equation of state and observing scenario, we compute the Bayes factor $B$ as a Savage-Dickey density ratio~\cite{Dickey1971}, the ratio of marginal posterior to marginal prior at $\Delta\Lambda = 0$, obtaining the former via a kernel density estimate. We plot the evolution of the Bayes factors in  Fig.~\ref{fig:bfs} as a function of the number of binary neutron star mergers detected within a redshift $z < 0.5$; the error bands quantify the variation in $B$ over 10 population realizations. As can be seen, for SKI5\_2009 the evidence is decidedly in favor of twin stars ($\log{B} < -6$) after a week of XG observations ($\sim 100$ detections). For SKI5\_2006, the evidence favors twin stars more modestly ($\log{B} \approx -3$), although some population realizations yield as decisive a result as for SKI5\_2009. For the other four cases, the Bayes factor test favors the no-twins hypothesis only moderately ($\log{B} \approx 3$) even after a month of XG observations ($\sim 500$ detections). This reinforces the conclusion that an XG network can confidently detect $\Delta\Lambda \gtrsim 100$, but is resolution-limited for smaller tidal deformability differences.

Having established the prospects for twin-star detection, we now investigate how well the twin-star parameters themselves can be measured. The main panels of Fig.~\ref{fig:results} relate the accuracy and precision with which $\Delta\Lambda$ and $M_t$ are recovered. 
We find that the recoveries are accurate at the 90\% confidence level for all of the equations of state we study. However, for SKI5\_2009, the mode of the posterior slightly overestimates the tidal deformability difference. We attribute this to the very small range $\Delta M_t \approx 0.01\Msun$ over which this equation of state supports twin stars, which means that the gaps in the binary tidal deformability vs chirp mass distribution are easily mimicked by statistical fluctuations in the data. For SKI272\_3505, the posterior mode instead slightly underestimates $M_t$, which we attribute to the proximity between $M_t$ and $M_{\rm TOV}$ for this equation of state.  
For the hadronic equations of state SKI5 and SK272, the $M_t$ parameter is meaningless, and we observe that the recovered value tends to be close to the midpoint of the neutron star mass spectrum. 

The size of the 90\% confidence regions in Fig.~\ref{fig:results} demonstrates the increasing precision in twin-star parameter measurements that can be expected from future gravitational-wave observations. Overall, we find that $M_t$ is more precisely constrained than $\Delta\Lambda$. In the best-case scenario, for SKI5\_2009, $M_t$ and $\Delta\Lambda$ are determined to within 1\% and 15\% uncertainty, respectively, after one month of observation. In contrast, after two years of A+ observations, the uncertainties are respectively 15\% and 68\%. Even in the most pessimistic scenario, for SK272\_3505, $M_t$ and $\Delta\Lambda$ are measured with respective errors of 20\% and 150\% after one month at XG sensitivity. Again, this illustrates the constraining power of XG observations.

\section{Discussion}

Our investigations demonstrate how hierarchical inference on a population of binary neutron star mergers can be used to systematically test for the presence of twin stars. We find that an XG network may be able to discern twin stars in the population after as little as one week of observation. Such a network can resolve a tidal deformability difference between twins of several hundred, and constrain the twin-star mass scale to within a few percent. When realized in future, measurements of this kind will guide the development of nuclear theory models for the neutron star equation of state.

While our analysis is not an exhaustive exploration of the parameter space for twin stars, we forecast parameter constraints for a variety of equations of state, including examples with first-order phase transitions of different onset densities and strengths. Besides the choice of equation of state, our forecasts also depend in principle on assumptions about the neutron star mass distribution and the relative abundance of hadronic vs hybrid twins in the detected population. We perform supplemental analyses in Appendix~\ref{sec:supp} that indicate that these additional factors do not strongly impact the recovery of the twin-star parameters.

Although the parametric $m$--$\Lambda$ relation model we adopt for the hierarchical inference is deliberately kept simple, our results indicate that it will suffice for accurate twin-star parameter measurements into the XG era. Nonetheless, when future observing campaigns return thousands of binary neutron star detections, it may be desirable to use a more sophisticated model that does not gloss over the finite extent of the twin-star mass range. One could also build the $m$--$\Lambda$ parameterization in terms of the fundamental properties of the phase transition itself, such as the onset density and the strength of the transition, in order to make measurements that are more directly informative for nuclear theory. The general hierarchical inference method laid out here can be adapted straightforwardly to accommodate these modifications. It would also work to detect other phenomena giving rise to ostensible twins, e.g.~an overlap in the neutron star and black hole mass spectra or a separate family of stellar-mass exotic compact objects.

\acknowledgments

The authors thank Katerina Chatziioannou and Jolien Creighton for useful suggestions about this work.
P.L. is supported by the Natural Sciences \& Engineering Research Council of Canada (NSERC).
K.C. is supported by the Czech Academy of Sciences under the project number LQ100102101.
The authors are grateful for computational resources provided by the LIGO Lab and supported by NSF Grants PHY-0757058 and PHY-0823459.

\bibliographystyle{apsrev4-1}
\bibliography{references}

\appendix

\section{Supplemental Material}\label{sec:supp}

In this appendix, we revisit some of the choices made in the main analysis and demonstrate that our conclusions are robust against alternative assumptions. We perform posterior predictive checks to reinforce the accuracy of our inferences. We also examine the connection between the $m$--$\Lambda$ relation for hybrid stars and the morphology of the binary tidal deformability vs chirp mass distribution in greater detail.

\subsection{Morphology of the tidal deformability distribution}

\begin{figure}[b]
    \centering
    \includegraphics[width=0.95\columnwidth,trim={0 0 30 30},clip]{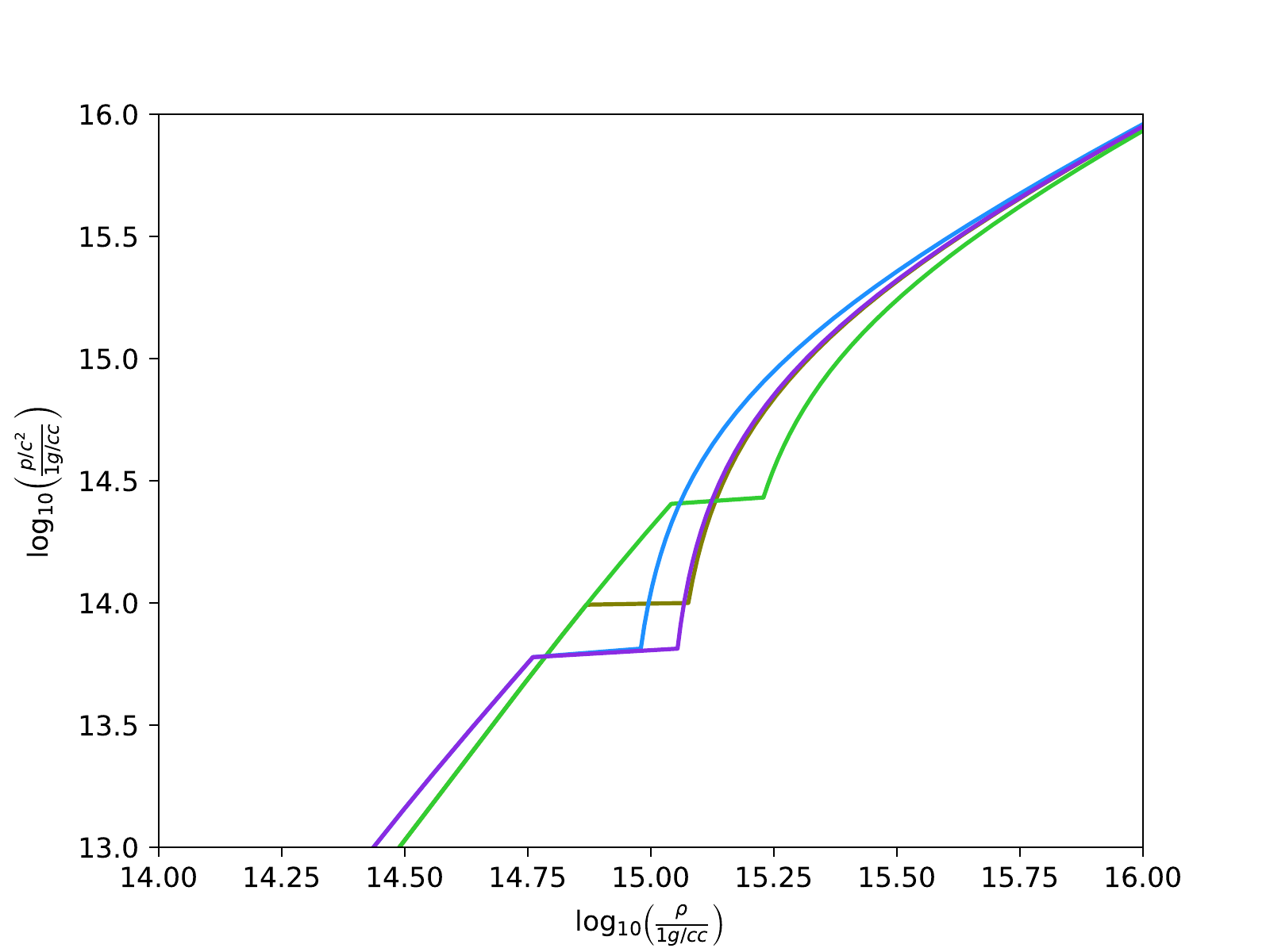}
    \caption{Twin-star--supporting equations of state used in this work. The discontinuities in baryon density at fixed pressure are first-order phase transitions of Maxwell type. The phase transitions occurring in these equations of state are strong enough to give rise to stable twin stars.}
    \label{fig:Eos_plot}
\end{figure}

\begin{figure}[t]
    \begin{tabular}{c}
      \includegraphics[width=0.95\columnwidth]{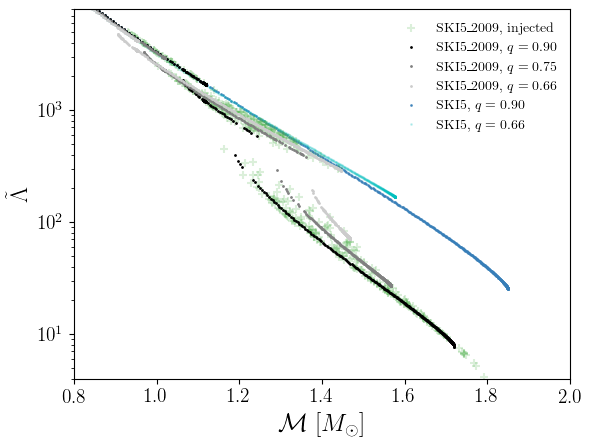} \\
      \includegraphics[width=0.95\columnwidth]{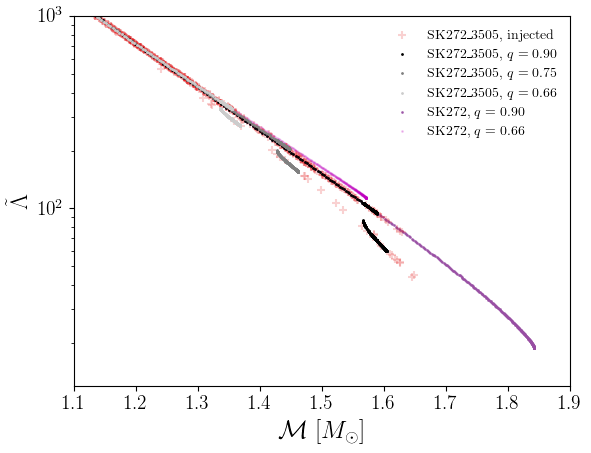}
    \end{tabular}
    \caption{The joint binary tidal deformability and chirp mass distribution probed by one month of XG observations for selected equations of state studied in this work. The distribution is broken down according to mass ratio, showing how the gaps that indicate the presence of twin stars arise. A uniform neutron star mass distribution is assumed.}
    \label{fig:lambdapop2}
\end{figure}

In the main analysis, we argued that an equation of state's support for twin stars translates into a disjoint two-dimensional distribution of binary tidal deformability

\begin{equation}
    \tilde{\Lambda} = \frac{16}{13} \frac{(1+12q) \Lambda_1 + (q+12)q^4\Lambda_2}{(1+q)^5}
\end{equation}
vs chirp mass

\begin{equation}
    \mathcal{M} = \frac{(m_1 m_2)^{3/5}}{(m_1+m_2)^{1/5}} ,
\end{equation}
whereas the equivalent distribution is contiguous for an equation of state without a disconnected hybrid star branch.
The gaps in the distribution originate directly from the unstable segment of the $m$--$\Lambda$ relation that connects the hadronic and hybrid branches for twin-star-supporting equations of state: these are the dashed segments in Fig.~\ref{fig:twins}.
They correspond to strong first-order phase transitions in the equation of state, where the baryon density jumps discontinuously at fixed pressure---see Fig.~\ref{fig:Eos_plot}.
The gaps in the $\tilde{\Lambda}$ vs $\mathcal{M}$ distribution occur where, along lines of constant mass ratio $q = m_2/m_1$, one component mass lies on the unstable segment.
This is illustrated in Fig.~\ref{fig:lambdapop2}, which breaks down the distribution in Fig.~\ref{fig:lambdapop} along lines of fixed mass ratio.
The upper (respectively, lower) gap in the distribution for SKI5\_2009 corresponds to $m_2$ ($m_1$) lying on the unstable segment. Note how a similar breakdown of the distribution for the purely hadronic SKI5 equation of state does not evince the appearance of any gaps.

Figure~\ref{fig:lambdapop2} also shows the same distribution for SK272\_3505 and its purely hadronic counterpart, SK272. In this case, due to its very small $\Delta M_t$ and $\Delta\Lambda$, the gaps in the distribution for SK272\_3505 are barely visible at scale. However, along lines of constant mass ratio, one can see the distribution is broken into distinct segments. Again, this contrasts with the continuous distribution for SK272.

\subsection{Posterior predictive checks}

\begin{figure*}
    \begin{tabular}{cc}
      \includegraphics[width=0.95\columnwidth]{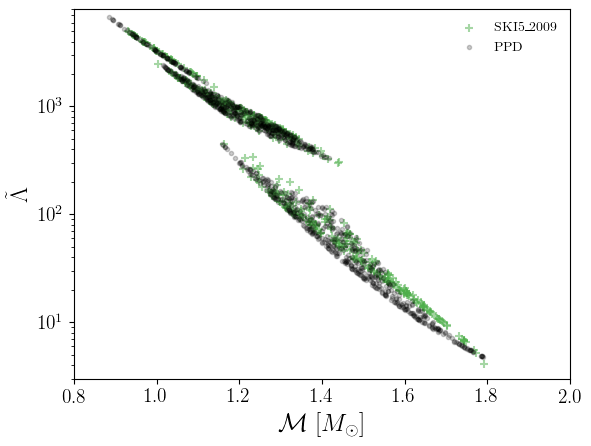} &
      \includegraphics[width=0.95\columnwidth]{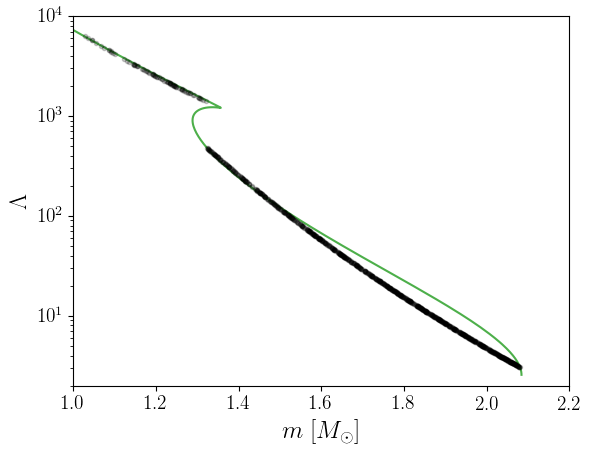} \\
      \includegraphics[width=0.95\columnwidth]{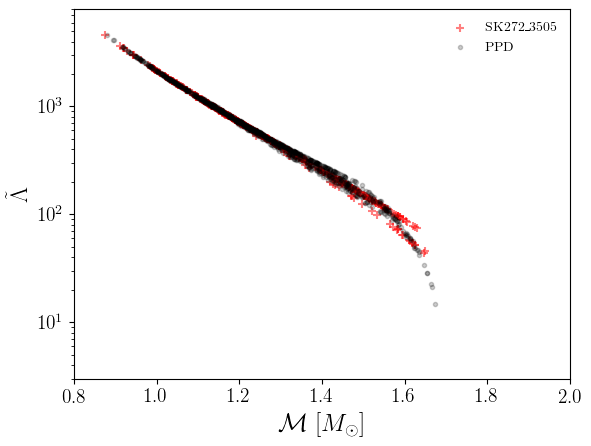} &
      \includegraphics[width=0.95\columnwidth]{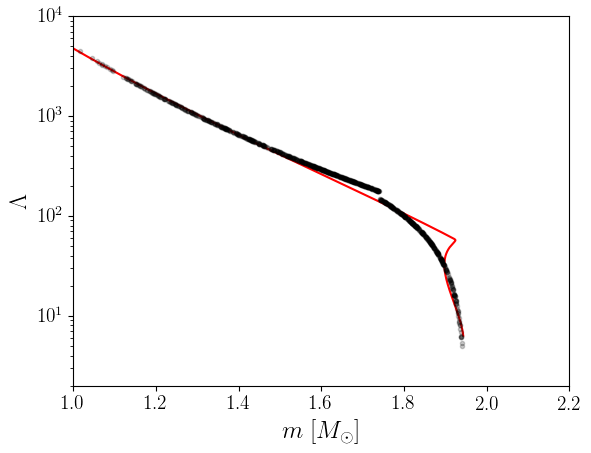}
    \end{tabular}
    \caption{Posterior predictive checks for selected twin-star-supporting equations of state. We compare the posterior predictive distribution (PPD) for the binary tidal deformability vs chirp mass distribution (left) and the mass--tidal deformability relation (right) to the injected values. The posterior predictive distribution is constructed from the approximate mode of the recovered three-dimensional posterior in $\Delta\Lambda$, $M_t$ and $k$.}
    \label{fig:ppd}
\end{figure*}

To bolster the results of the main analysis, we perform posterior predictive checks and examine how constraints on the twin-star parameters evolve with the number of binary neutron star detections. For each injected equation of state, we extract the approximate maximum-likelihood parameters ($\Delta\Lambda$,$M_t$,$k$) from the population-averaged posterior after one month of XG observations. We use these parameters to reconstruct the best-fit $m$--$\Lambda$ relation according to Eq.~\eqref{fit}. We then sample a uniform binary neutron star population and reconstruct the predicted distribution of binary tidal deformability vs chirp mass for the population. Figure~\ref{fig:ppd} compares these predictions for the $m$--$\Lambda$ relation and the $\tilde{\Lambda}$ vs $\mathcal{M}$ distribution to their true values for two selected equations of state, SKI5\_2009 and SK272\_3505.

For SKI5\_2009, we observe that the predicted distributions are close, but not perfect, matches to the injected ones. The main discrepancy occurs at large masses, where the model's fixed power-law slope does not allow for the flexibility required to track the actual $m$ vs $\Lambda$ trend. This is essentially a deliberate tradeoff in the model, which prizes simplicity over fidelity. Similarly, by construction, the model does not track the $m$--$\Lambda$ relation right through the unstable segment connecting the hadronic and hybrid branches. However, our simplified treatment of this juncture manifestly reproduces the right $\tilde{\Lambda}$ vs $\mathcal{M}$ morphology and accurately locates the discontinuity in the $m$--$\Lambda$ relation.

For SK272\_3505, which has a much higher phase transition onset density and much smaller $\Delta M_t$ and $\Delta\Lambda$, the model clearly struggles to match both the location and extent of the hybrid branch. The mode of the posterior, upon which the illustrated posterior predictive check is based, favors a good fit to the latter at the expense of the former. However, as this equation of state is the one with the largest uncertainties \textit{a posteriori}, the posterior encompasses other parameter combinations that recover the twin star mass scale more accurately.

\begin{figure*}
    \begin{tabular}{cc}
      \includegraphics[width=0.45\textwidth]{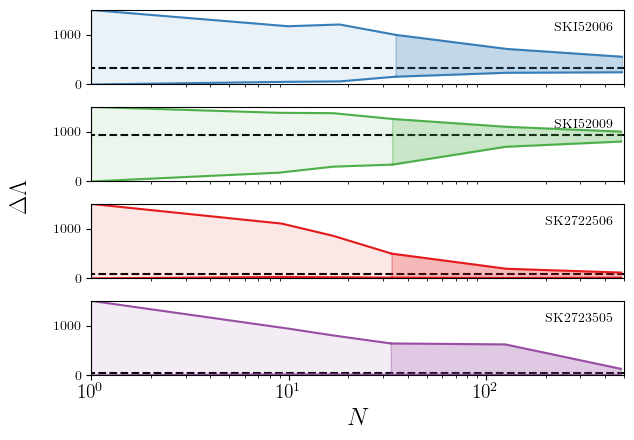}   & \includegraphics[width=0.45\textwidth]{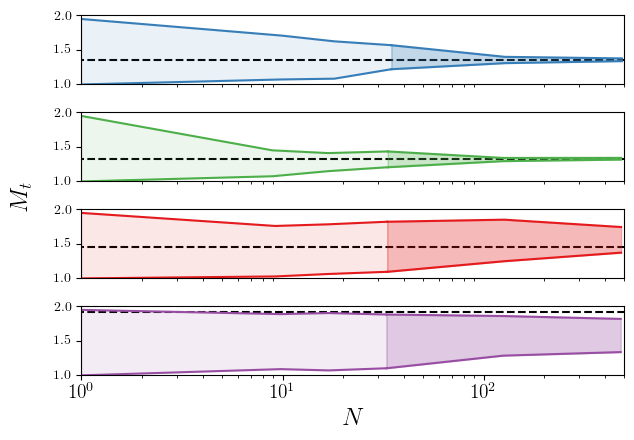}
    \end{tabular}
    \caption{Constraints on the tidal deformability difference between twins and the twin-star mass scale as a function of the number of binary neutron star mergers detected within a redshift $z < 0.5$. We show the evolution of the 90\% confidence contours. The era of A+ (respectively, XG) observations is shown with lighter (darker) shading. We average over 10 population realizations to mitigate fluctuations in statistical uncertainties.}
    \label{fig:params}
\end{figure*}

This can be seen in the evolution of the twin star parameter uncertainties as a function of the number of $z < 0.5$ binary neutron star detections. In Fig.~\ref{fig:params}, we show the posterior 90\% confidence intervals on $\Delta\Lambda$ and $M_t$ for the scenarios that are presented in Fig.~\ref{fig:results}. The parameters are recovered accurately for all of the equations of state, except for a slight underestimate of $M_t$ in the case of SK272\_3505, which is related to the mismatch in the posterior predictive check. The precision in the parameter measurements increases significantly when passing from the A+ to the XG detector network.

\subsection{Effect of the population realization}

\begin{figure}[t]
      \includegraphics[width=0.95\columnwidth]{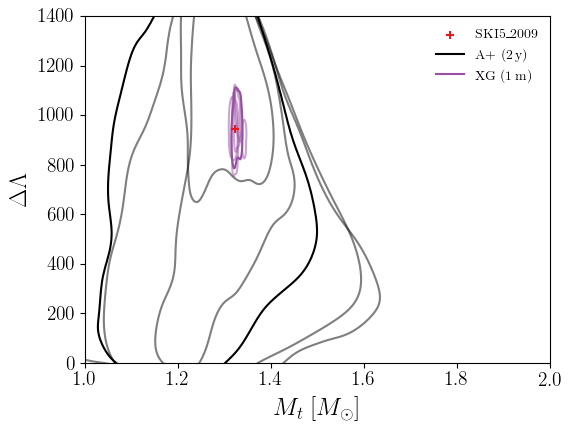}
    \caption{Posteriors on twin-star parameters of the mass-tidal deformability relation for the SKI5\_2009 equation of state. The 90\% credible region of the posterior is shown for the best-case A+ and XG observing scenarios we consider. We show how the credible regions differ across several different population realizations (faint traces) compared to the population averages (solid lines) that were displayed in the equivalent panel of Fig.~\ref{fig:results}. Note how the variation due to the population realization diminishes as more observations are accumulated.}
    \label{fig:results_indiv}
\end{figure}

The results presented in the main analysis average over 10 realizations of the binary neutron star population. Here we show how much the parameter constraints vary with the population realization. In Fig.~\ref{fig:results_indiv}, we show the two-dimensional posterior 90\% confidence contours on $\Delta\Lambda$ and $M_t$ for SKI5\_2009 for five population realizations after two years of A+ observations and one month of XG observations---these are the best-case A+ and XG observing scenarios, respectively. One can see that statistical fluctuations due to the population realization can significantly impact the A+ contours; however, as the detector network improves and the number of observations increases, this variability becomes less significant.

\subsection{Effect of the branching ratio}

\begin{figure}[t]
    \includegraphics[width=0.95\columnwidth]{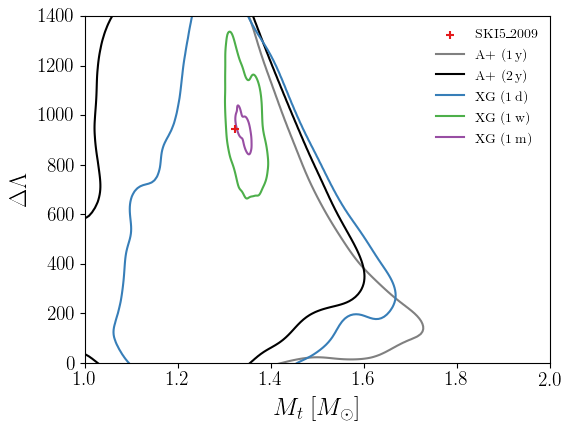}
    \caption{Posteriors on twin-star parameters of the mass-tidal deformability relation for the SKI5\_2009 equation of state, under a different assumption about the hadronic vs hybrid branching ratio for twin stars. The 90\% credible region of the posterior, averaged over 10 population realizations, is shown for various observing scenarios. Note the qualitative similarity with the equivalent panel of Fig.~\ref{fig:results}.}
    \label{fig:branch_results}
\end{figure}

We assumed above that neutron stars with masses in the twin star range lie with equal probability on the hadronic or hybrid branch of the $m$--$\Lambda$ relation. Here we make an alternative choice and show that the twin-star parameter constraints are largely unchanged in our best-case scenario.  We assume that twin stars are distributed uniformly in central density, such that the probability for an arbitrary twin star to lie on the hadronic vs the hybrid branch is proportional to the ratio of density ranges supported by each twin-star branch. For instance, if hadronic twins span a central density range of $0.3\,\rho_{\rm nuc}$, and hybrid twins span only $0.1\,\rho_{\rm nuc}$, there should be three times as many hadronic twins as hybrid ones. Repeating the main analysis for the SKI5\_2009 equation of state under these conditions, we obtain Fig.~\ref{fig:bimod_results}'s population-averaged posterior on the parameters $\Delta\Lambda$ and $M_t$. As was the case under the original branching ratio assumption, twin stars are definitively identified in the population after one week of XG observations.

\subsection{Effect of the mass distribution}

\begin{figure*}
    \centering
    \includegraphics[width=0.95\textwidth,trim={145 160 130 70},clip]{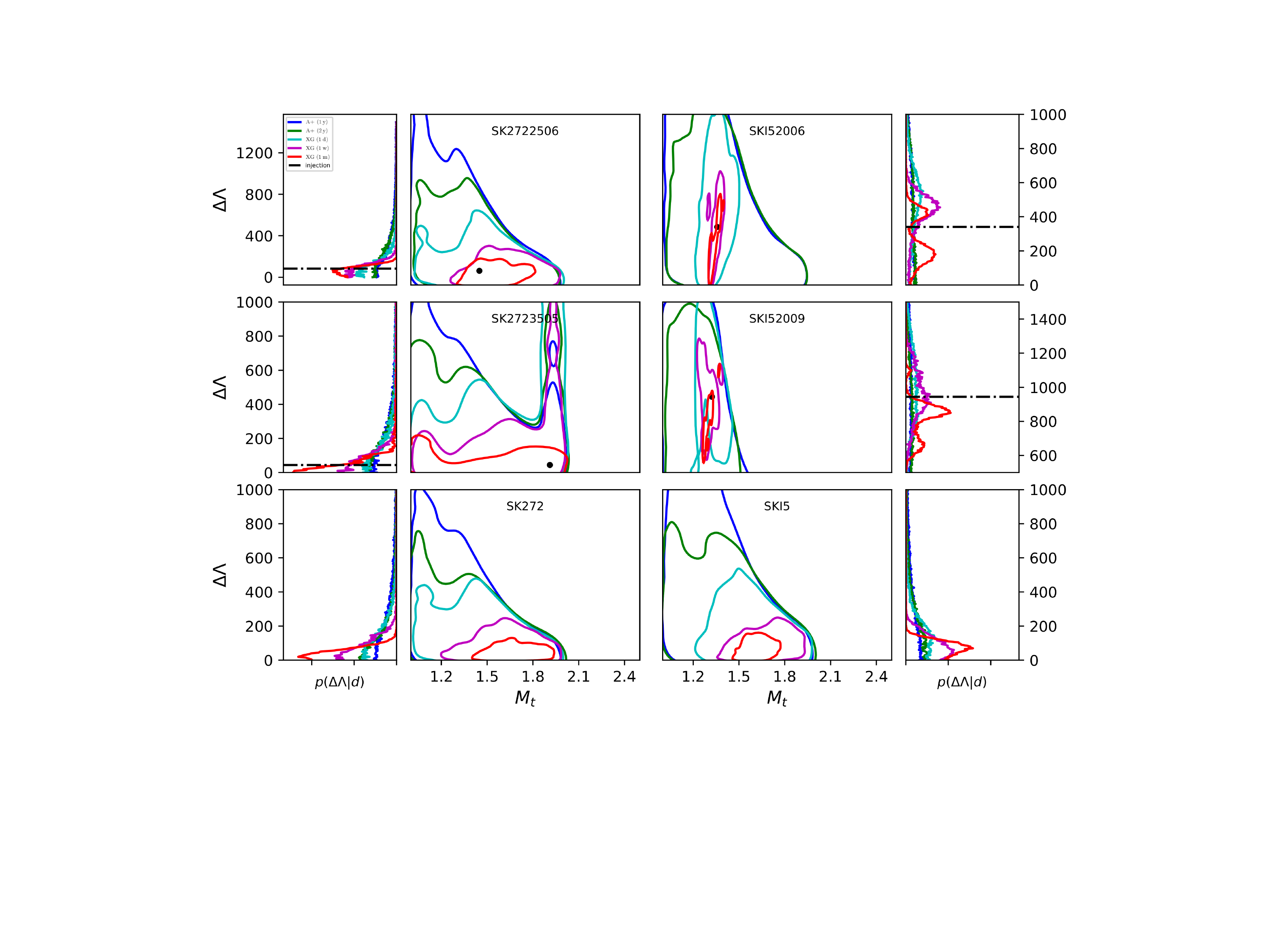}
    \caption{Posteriors on twin-star parameters of the mass-tidal deformability relation for different equations of state, under the assumption of a bimodal neutron star mass distribution. The 90\% credible region of the posterior, averaged over 10 population realizations, is shown for various observing scenarios. Note the qualitative similarity with Fig.~\ref{fig:results}.}
    \label{fig:bimod_results}
\end{figure*}

In the main analysis, we adopted a uniform mass distribution for the binary neutron star population.
We now revisit that assumption and demonstrate that our conclusions are unchanged if we instead use the bimodal neutron star mass distribution from Ref.~\cite{FarrChatziioannou2020}, inspired by observations of Galactic pulsars. We continue to assume that neutron stars pair randomly into binaries.
When we repeat the main analysis with this new population model, we obtain the twin-star parameter constraints shown in Fig.~\ref{fig:bimod_results}.
As above, the results are averaged over 10 population realizations.
We obtain qualitatively similar constraints as with the uniform neutron star mass distribution: one month of XG observations can identify the presence of twin stars in the SKI5\_2009 and SKI5\_2006 scenarios, while $\Delta\Lambda \gtrsim 100$ is ruled out in the other four scenarios.

\end{document}